\documentclass[sigconf]{acmart}
\usepackage{cleveref}
\usepackage{bm}
\usepackage{booktabs,subcaption,amsfonts,dcolumn}
\usepackage{wasysym}
\usepackage{commath}

\DeclareMathOperator{\E}{\mathbb{E}}
 \usepackage{multirow}
 \usepackage[ruled,vlined,linesnumbered]{algorithm2e}

\AtBeginDocument{%
  \providecommand\BibTeX{{%
    \normalfont B\kern-0.5em{\scshape i\kern-0.25em b}\kern-0.8em\TeX}}}

\setcopyright{iw3c2w3} 
\copyrightyear{2021}
\acmYear{2021} 
\acmConference[WWW '21]{Proceedings of the Web Conference 2021}{April 19--23, 2021}{Ljubljana, Slovenia} 
\acmBooktitle{Proceedings of the Web Conference 2021 (WWW '21), April 19--23, 2021, Ljubljana, Slovenia}
\acmPrice{}
\acmDOI{10.1145/3442381.3449901}
\acmISBN{978-1-4503-8312-7/21/04}
\begin{document}

\title{Maximizing Marginal Fairness for Dynamic Learning to Rank}

\author{Tao Yang}

\affiliation{%
   \institution{University of Utah}
     \streetaddress{50 Central Campus Dr.}
   \city{Salt Lake City}
   \state{Utah}
   \postcode{84112}
}
\email{taoyang@cs.utah.edu}

\author{Qingyao Ai}
\affiliation{%
   \institution{University of Utah}
     \streetaddress{50 Central Campus Dr.}
   \city{ Salt Lake City}
   \state{Utah}
   \postcode{84112}
}
\email{aiqy@cs.utah.edu}


\begin{abstract}
Rankings, especially those in search and recommendation systems, often determine how people access information and how information is exposed to people.
Therefore, how to balance the relevance and fairness of information exposure is considered as one of the key problems for modern IR systems.
As conventional ranking frameworks that myopically sorts documents with their relevance will inevitably introduce unfair result exposure, recent studies on ranking fairness mostly focus on dynamic ranking paradigms where result rankings can be adapted in real-time to support fairness in groups (i.e., races, genders, etc.).
Existing studies on fairness in dynamic learning to rank, however, 
often achieve the overall fairness of document exposure in ranked lists by significantly sacrificing the performance of result relevance and fairness on the top results.
To address this problem, we propose a fair and unbiased ranking method named Maximal Marginal Fairness (MMF). 
The algorithm integrates unbiased estimators for both relevance and merit-based fairness while providing an explicit controller that balances the selection of documents to maximize the marginal relevance and fairness in top-k results.
Theoretical and empirical analysis shows that, with small compromises on
long list fairness, our method achieves superior efficiency and effectiveness comparing to the state-of-the-art algorithms in both relevance and fairness for top-k rankings.

\end{abstract}




\maketitle
\section{Introduction}

Fairness in ranking has drawn much attention as ranking systems, especially those in search and recommendation systems, could significantly affect how people access information and how information is exposed to users \cite{biega2018equity}. 
For example, job posts ranked highly on LinkedIn are more likely to receive more applications; new products on the bottom of Amazon search result pages are less likely to be clicked by customers.
Without proper treatments, ranking systems could introduce unintended unfairness to various aspects of people's lives, such  as job opportunities, economical gain, etc.

As traditional ranking algorithms that produce static ranked lists can hardly handle the unfairness of result exposure in practice, the ranking paradigms that dynamically change result ranking on the fly, namely the dynamic Learning-to-Rank (LTR) \cite{morik2020controlling}, have received more and more attention in the research communities.
The idea of dynamic LTR is to learn and adapt ranking models based on user feedback in real time so that past user interactions or result distributions could affect the exposure of future results.
Through heavily exposed to implicit user examination bias \cite{joachims2005accurately, joachims2007evaluating, yue2010beyond} and selection bias \cite{wang2016learning,ovaisi2020correcting}, dynamic LTR allows ranking systems to produce multiple ranked lists for a single request, which makes it possible to explicitly control or balance the exposure of results in different groups (i.e., race, gender, etc.) for ranking fairness.

Existing studies on ranking fairness in dynamic LTR mostly focuses on developing effective algorithms to achieve merit-based fairness \cite{biega2018equity,singh2018fairness}.
Well-known examples include the linear programming algorithm (LinProg) \cite{singh2018fairness} that determines result ranking by taking group fairness as an optimization constraint, and the FairCo algorithm \cite{morik2020controlling} that manipulates ranking scores dynamically according to the current exposure of results in different groups.
Despite their solid theoretical foundations, these algorithms often sacrifice the performance of result relevance significantly on the top results of each ranked list.
These trade-offs are uncontrollable as they are difficult to be quantified explicitly, which makes the risk of applying fairness algorithms in practice still high today.

In this paper, we present a novel fairness algorithm for dynamic LTR that considers both relevance and fairness for ranking criterion.
We believe that it is important to control and balance result relevance and fairness in online ranking systems, especially for the top-k results in ranked lists.
Therefore, inspired by the studies of search diversification that try to improve novelty while preserving ranking performance (especially the maximal marginal relevance frameworks \cite{carbonell1998use}), we propose a Maximal Marginal Fairness (MMF) algorithm that optimizes ranking performance while explicitly mitigating amortized group unfairness for selected items.
MMF dynamically selects documents that are relevant and underexposed in top-k results to maximize the marginal relevance and fairness.
With a small compromise on the bottom of long ranked lists, MMF achieves superior performance and significantly outperforms the state-of-the-art fairness algorithms not only in top-k relevance, but also in top-k fairness.
As most people would examine or only examine the top results on a result page \cite{joachims2002optimizing,craswell2008experimental,wang2013incorporating,Malkevich:2017:EAC:3121050.3121096}, MMF is highly competitive and preferable in real LTR systems and applications.

From technical viewpoints, the main contribution of this paper are two-fold. 
First, extending the existing definition of merit-based group fairness in ranking \cite{biega2018equity,singh2018fairness}, we develop a metric to measure the group fairness of exposure in the top-k rankings. 
We show that most existing state-of-the-art methods for ranking fairness focus more on the overall fairness of document exposure while compromising a lot in the top ranks of each ranked list. 
Second, we propose a Maximal Marginal Fairness (MMF) algorithm that can explicitly control and balance result relevance and fairness in top-k rankings. 
In particular, our method uses a hyper-parameter $\lambda$ to determine whether the system needs to select an item being relevant or an item that improves the marginal fairness of the current ranked list.
We evaluate and compare our algorithm with existing state-of-the-art methods on both synthetic and real-world preference datasets using simulated user interactions.
Theoretical and empirical analysis shows that our method can achieve significant efficiency and effectiveness improvements in top-k relevance and fairness. 

\vspace{-0.3cm}
\section{Related work}

Leveraging biased click data for optimizing learning to rank systems has been a popular approach in information retrieval \cite{joachims2002optimizing,joachims2005accurately}. 
As click data are often noisy and biased, numerous unbiased learning to rank (ULTR) methods have been proposed based on different theoretical foundations  \cite{ai2020unbiased}, including click models \cite{jin2020deep,wang2018position,dupret2008user}, randomization \cite{radlinski2006minimally}, causal inference \cite{ai2018unbiased,agarwal2019general,joachims2017unbiased}, etc. 
Those methods make it possible to achieve unbiased relevance estimation or rankings for learning to rank in noisy environments.


Ranking according to intrinsic result relevance is important yet not enough today. 
One of the key principle \cite{robertson1977probability} of ranking in Information Retrieval states that documents should be ranked in order of the probability of relevance or usefulness. 
Such arguments, however, only consider the responsibilities of ranking systems to users while ignoring the items being ranked. 
Recently, there has been a growing concern about fairness behind ranking algorithms in both academia and industry \cite{zehlike2017fa,biega2018equity,zehlike2020reducing}. 
Since rankings are the main interface from which we find content, products, music, news and etc., their ordering contributes not only to the utility of users, but also to information providers.

To address this, some studies focus on restricting the fraction of items of each attribute in a ranking \cite{zehlike2017fa,yang2017measuring}.
While another natural way of understanding unfairness is by considering difference in exposure, which directly relates unfairness to economic or social opportunities \cite{zehlike2020reducing}. 
For example, researchers try to achieve amortized fairness of exposure by reducing group disparate exposure for only the top position  \cite{zehlike2020reducing} or making exposure proportional to relevance \cite{morik2020controlling,singh2019policy,singh2018fairness,biega2018equity}. 
The reason for reducing exposure disparity is that minimal difference in relevance can result in a large difference in exposure across groups  \cite{singh2018fairness} if the ranking only is based on relevance, since there might exist a large skew in the distribution of exposure like position bias \cite{joachims2017accurately}, i.e., users observe way much more top ranks than the bottom part. 
To achieve this, various methods have been proposed. 
For example, fairness loss based on Softmax function is proposed by \citet{zehlike2020reducing}, which considers equal group exposure in top rank. Linear programming methods are proposed in  \cite{singh2018fairness,biega2018equity,celis2018ranking}, which try to give a general framework for computing optimal probabilistic rankings for merit-based fairness. 
The policy learning method is adopted to maximize ranking metrics while introducing unfairness as additional loss in  \cite{singh2019policy}. 

In fact, the ideas behind existing fairness methods are always to avoid showing items from the same class or group, which is not a new topic and have already been studied as diversity and novelty in the information Retrieval Community for decades \cite{xia2015learning,xia2017adapting,jiang2017learning}. 
Though the utility of diversified ranking is still centered on users not for providers, studies on search diversification also focus on optimizing result ranking beyond information relevance. 
One of the most famous approaches is the maximal marginal relevance approach (MMR)  \cite{carbonell1998use} that treats ranking as a Markov Decision Process by selecting documents that maximize the combination of sub-topic relevance given previously selected results.

In a recent work  \cite{morik2020controlling}, the proportional controller in control theory is applied to mitigate unfairness in dynamic LTR settings. 
However, existing fairness algorithms in dynamic LTR only consider overall fairness while ignoring the top-k unfairness. 
In this paper, we first define the top-k group fairness, which is often ignored by existing works. 
Then, based on the top-k fairness metrics, inspired by MMR, we propose a concept named marginal fairness and directly optimize top-k relevance and fairness in a dynamic LTR environment where both the relevance and fairness modules of the algorithms are learned and adapted according to real-time user feedback.

Another algorithm is related with our work is FA*IR, proposed by \citet{zehlike2017fa}.
FA*IR is a post-processing method to explicitly guarantees the exposure of documents in top-k ranking. The key difference between our algorithm MMF and FA*IR is that FA*IR focuses on a simplified offline scenario and ignore the fact that result exposure could affect relevance estimations in LTR.
\vspace{-0.4cm}
\section{PROBLEM FORMULATION}
In this section, we introduce the problem of relevance and fairness estimation in dynamic LTR with a focus on the top-k results.
A summary of the notations used in this paper is shown in Table \ref{tab:notation}.

\begin{table}[t]
	\setlength{\belowcaptionskip}{-10pt}
	\caption{A summary of notations.}
	 \vspace{-0.4cm}
	\small
	\def\arraystretch{1}
	\begin{tabular}
		{| p{0.07\textwidth} | p{0.36\textwidth}|} \hline
		
		$\bm{\sigma_t}$,$\bm{x_t}$,$\bm{c_t}$, 
		$\bm{o_t}$, $\bm{p_t}$, $\bm{r_t}$ & The presented document ranking ($\bm{\sigma_t}$), the corresponding feature vectors ($\bm{x_t}$), the clicks on each document ($\bm{c_t}$), the binary variables indicating user's examination on each document ($\bm{o_t}$), user's propensity to examine results on certain positions ($\bm{p_t}$, namely $P(\bm{o_t}=1)$), and the true (personalized) binary relevance rating of the documents ($\bm{r_t}$) at time step $t$ in dynamic LTR setting.   \\\hline
		$R(d)$ &  The average relevance across all users for document $d$.\\\hline
		$R_\theta (d|\bm{x_t})$ & The predicted relevance of $d$ given by the model parameterized by  
		$\theta$.\\\hline
		$\mathcal{G}$,$G_i$,$\mathbb{G}_i^k$ & The set of groups to consider ($\mathcal{G}$) and the $i\-th$ group ($G_i$). $\mathbb{G}_i^k$ is the priority queue of size k for $G_i$ according to estimated relevance.\\\hline

	\end{tabular}\label{tab:notation}
	\vspace{-0.5cm}
\end{table}

\vspace{-0.4cm}
\subsection{Relevance Estimation in Dynamic LTR}

In dynamic LTR frameworks, the most popular paradigm for relevance estimation is to infer relevance from user feedback directly.
Specifically, from partial and biased feedback such as user clicks, we need to construct a model to estimate relevance. 
In general, traditional LTR models can be categorized as cardinal and ordinal ones. 
Ordinal LTR models give ordinal numbers according to output scores for items, while those scores themselves have no exact meaning. 
Cardinal LTR models, on the other hand, predict document ranking scores that are proportional or directly reflect the relevance of the documents. 
Since exposure disparity explicitly involves relevance, similar to previous studies \cite{morik2020controlling}, we only focus on cardinal LTR models in this paper. 
Here, we adopt a LTR model $R_\theta (d|\bm{x_t})$ parameterized by $\theta$ with least-square loss as
\begin{equation}
\begin{split}
\mathcal{L}^\tau(\theta)&=\sum_{t=1}^\tau\sum_d\bigg(\bm{r_t}(d)-R_\theta(d|\bm{x_t})\bigg)^2   \\
&\stackrel{\Delta}{=}\sum_{t=1}^\tau\sum_d\bigg(R_\theta(d|\bm{x_t})^2-2\bm{r_t}(d)R_\theta(d|\bm{x_t})\bigg)\\
\end{split}
\label{eq:skyline}
\end{equation}
where $\tau$ is the total number of existing time steps in dynamic LTR, and $\stackrel{\Delta}{=}$ means equal while ignoring constants.

As the true relevance judgements $\bm{r_t}$ are not available, based on the studies of unbiased learning to rank \cite{joachims2017unbiased}, we define an unbiased estimation of $\mathcal{L}^\tau(\theta)$ using click data $\bm{c_t}$ by applying Inverse Propensity Score (IPS) weighting as 
\begin{equation}
\widetilde{\mathcal{L}}^\tau(\theta)=\sum_{t=1}^\tau\sum_d(R_\theta(d|\mathbf{\mathbf{\bm{x_t}}})^2-2\frac{\mathbf{c_t(d)}}{\bm{p_t}(d)}R_\theta(d|\mathbf{\mathbf{\bm{x_t}}}))
\label{eq:IPS_loss}
\end{equation}
where $\bm{p_t}$ is the user's examination propensities on each result position (namely $P(\bm{o_t}=1)$), and we assume that
\begin{equation}
\begin{split}
P(c=1)&= P(o=1)\cdot P(r=1)
\end{split}
\label{eq:click}
\end{equation}
which means that users click a search result ($c=1$) only when it is both
observed ($o=1$) and perceived as relevant ($r=1$), and $o$ and $r$ are independent.
The unbiasness of $\widetilde{\mathcal{L}}^\tau(\theta)$ can be proven as below
\begin{equation}
\begin{split}
\E_{\mathbf{\bm{o_t}}}[\widetilde{\mathcal{L}}^\tau(\theta)]&=\sum_{t=1}^\tau\sum_d(R_\theta(d|\mathbf{\bm{x_t}})^2-2\frac{\E_{\mathbf{\mathbf{\bm{o_t}}}}[\mathbf{c_t(d)}]}{\bm{p_t}(d)}R_\theta(d|\mathbf{\bm{x_t}}))\\
&=\sum_{t=1}^\tau\sum_d(R_\theta(d|\mathbf{\bm{x_t}})^2-2\frac{\bm{p_t}(d)\mathbf{\bm{r_t}(d)}}{\bm{p_t}(d)}R_\theta(d|\mathbf{\bm{x_t}}))\\
&=\sum_{t=1}^\tau\sum_d(R_\theta(d|\mathbf{\bm{x_t}})^2-2\mathbf{\bm{r_t}(d)}R_\theta(d|\mathbf{\bm{x_t}})) = \mathcal{L}^\tau(\theta)\\
\end{split}
\label{eq:unbias}
\end{equation}
Similarly, we could get unbiased estimation of the average relevance of $d$ across all users ($R(d)$) as 
\begin{equation}
\begin{split}
R^{IPS}(d)=\frac{1}{\tau}\sum_{t=1}^\tau\frac{c_t(d)}{p_t(d)}
\end{split}
\label{eq:average_relevance}
\end{equation}
$R^{IPS}(d)$ is used for both fairness controlling and as a global ranking baseline without personalization (i.e., considering document relevance with respect to each individual user). 
The estimation of position bias $p_t(d)$ can be achieved through various methods in advance \cite{ai2018unbiased,wang2018position,agarwal2019estimating}, which is not in the scope of this paper.

\subsection{Merit-based Top-k Fairness}
We now introduce the definition of the Top-k merit-based group unfairness in dynamic LTR. 
First, similar to previous studies \cite{biega2018equity, singh2018fairness, morik2020controlling}, we define the exposure of a document $d$ as its marginal probability of being examined $\bm{p_t}(d)=P(\bm{o_t}(d)=1|\bm{\sigma_t},\bm{x_t},\bm{r_t})$ where $\bm{\sigma_t},\bm{x_t},\bm{r_t}$ are the presented ranking, feature vectors, and true relevance of documents at time step $t$ in dynamic LTR. 
Let $\mathcal{G}=\{G_1,...,G_m\}$ be the possible groups that each document could belong to.
Suppose that we only care about the exposure fairness in top-k positions, then we can define group-based fairness in top-k positions following the merit-based fairness definition \cite{ morik2020controlling} as
\begin{equation}
    Exp_t^k(G_i)=\frac{1}{| G_i 
    |}\sum_{d\in G_i \cap  d\in \bm{\sigma_t}^k}p_t(d)
    \label{eq:exp}
\end{equation}
where $\bm{\sigma_t}^k$ is the top k documents in presented ranking $\bm{\sigma_t}$, $\bm{p_t}(d)$ is the examination propensity on $d$ at time step $t$.
Following previous studies, we define the merit of a group or document as the expected average relevance across all documents in the same group:
\begin{equation}
    Merit(G_i)=\frac{1}{|G_i|}\sum_{d\in G_i}R(d)
\end{equation}
Thus, for any two groups $G_i$ and $G_j$, we can define their accumulative disparity in top-k positions as
\begin{equation}
\begin{split}
Exp\_cum_{\tau}^k(G_i)&=\sum_{t=1}^\tau Exp_t^k(G_i)\\
Exp\_Mer_\tau^k(G_i)&=\frac{\frac{1}{\tau}Exp\_cum_{\tau}^k(G_i)}{Merit(G_i)}\\
D_{\tau}^k(G_i,G_j)&=\norm{Exp\_Mer_\tau^k(G_i)-Exp\_Mer_\tau^k(G_j)}
\label{eq:pair_disparity}
\end{split}
\end{equation}
Thus, larger disparity indicates greater violation of fairness in top-k rankings.
Additional to pairwise disparity defined in Eq.\ref{eq:pair_disparity}, for ranking with more than 2 groups, we
define the unfairness of a list as the average disparity of all pairs:
\begin{equation}
    \text{Unfairness@k} = \overline{D_{\tau}^k}=\frac{2}{m(m-1)}\sum_{i=0}^m\sum_{j=i+1}^m\norm{D_{\tau}^k(G_i,G_j)}
\end{equation}
Note that the original merit-based unfairness defined in  \cite{morik2020controlling} can be seen as a special case in our formulation where $k$ is the number of all possible documents.
In this paper, our goal is to create rankings with high relevance $\bm{r_\tau}$ while maintaining a low unfairness $\overline{D_{\tau}^k}$ in top-k results after the time step $\tau$ in dynamic LTR.

\section{Our Approach}

In this section, we describe our approach to balance relevance and fairness in dynamic LTR. 
Specifically, we first introduce the concept of maximal marginal fairness, then we discuss our MMF algorithm.

\subsection{Maximal Marginal Fairness}

In this paper, we define a new concept for ranking fairness as \textit{Marginal Fairness}.
Ranking can be modeled as a greedy selection process where we create a ranked list by sequentially selecting documents from a candidate pool.
Based on this assumption, considerable ranking algorithms have been proposed to optimize ranking utility from various perspectives such as information relevance \cite{liu2011learning}, novelty \cite{rodrygo2015search}, etc.
Particularly, in the studies of search diversification, one of the most well-known algorithms is the Maximal Marginal Relevance (MMR) algorithm \cite{carbonell1998use} that greedily selects documents based on their \textit{Marginal Relevance}, the maximal utility of a document given the selected results in the current ranked list, to balance ranking relevance and novelty. 
Inspired by MMR and the concept of marginal relevance, we model ranking as a greedy selection problem and define marginal fairness as \textit{the marginal gain of fairness, or the marginal reduction of unfairness, when selecting and adding a document given the selected results in the current ranked list}.

$d_\tau^k$ is the document we select for $k^{th}$ position in the ranked list $\bm{\sigma_\tau}$.
Formally,
then the marginal fairness of selecting document $d_\tau^k$ from group $G$ is  
\begin{equation}
MF(G|\sigma_\tau^{k-1}) =\overline{D_{\tau}^{k-1}} - \overline{D_{\tau}^k}(d_\tau^k), \;\text{where}\; d_\tau^k\in G
\label{equ:marginal_fairness}
\end{equation}
where $\overline{D_{\tau}^k}(d_\tau^k)$ is the unfairness after we select $d_\tau^k$.

Then, to maximize merit-based fairness of top $k$ results, a straightforward method is to maximize marginal fairness by selecting the document with maximum $MF(G|\sigma_\tau^{k-1})$. 
This observation serves as our foundation for the construction of the MMF algorithm.

\subsection{Algorithm}

The MMF algorithm includes three sub-modules -- the selection of documents for maximizing marginal fairness, the selection of documents for maximizing relevance, and the controller that balances top-k relevance and fairness.

\subsubsection{Fairness Module}
As discussed previously, MMF optimizes top-k fairness through greedily selecting documents to maximize marginal fairness.
In Eq.~(\ref{equ:marginal_fairness}), the computation of marginal fairness  requires us to compute the current and updated disparities ($D_{\tau}^{k-1}(G_i,G_j)$ and $D_{\tau}^{k}(G_i,G_j)$) for every pair of groups.
In practice, however, we can prove that maximal marginal fairness can be achieved directly by selecting a document from the group with lowest expected merit $Exp\_Mer_\tau^k(G_i)$ in Eq.~(\ref{eq:pair_disparity}) given that the scale of $\bm{p_\tau}(d)$ is much smaller than $Exp\_Mer_\tau^k(G_i)$\footnote{We ignore the proof for simplicity}.
Also, as the true relevance of documents $R(d)$ is not available, we use $R^{IPS}(d)$ to estimate $R(d)$ and get an unbiased estimation of $Exp\_Mer_\tau^k(G_i)$ as $\hat{Exp\_Mer_\tau^k(G_i)}$.

Thus, we compute the best group to select documents from MMF as
\begin{equation}
\begin{split}
G_{\tau}^k&=\underset{G}{max}~MF(G|\sigma_\tau^{k-1})\ =  \underset{G}{min}~\hat{Exp\_Mer_\tau^{k}(G)}
\end{split}
\label{eq:group_select}
\end{equation}
Note that every document in $G_{\tau}^k$ would have the maximal marginal fairness given the ranked list $\sigma_\tau^{k-1}$.



\subsubsection{Relevance Module}
To maximize the performance of top-k results in terms of ranking relevance, the optimal solution is to select documents based on their estimated relevance from user interactions (i.e., $R^{IPS}$) or a LTR model parameterized by $\theta$ ($R^{\theta}$) as
\begin{equation}
\begin{split}
\bar{d_{\tau}^k}&=\underset{ d \not\in \sigma_\tau^{k-1}}{argmax}\; R^{\theta}(d)
\end{split}
\label{eq:relevance_select}
\end{equation}
where we only consider relevance but not fairness in selecting the next document for the ranked list.
In case we need to select documents from the group with maximal marginal fairness, we could apply similar process and get
\begin{equation}
\begin{split}
\Tilde{d_{\tau}^k}&=\underset{d\in G_{\tau}^k \land d \not\in \sigma_\tau^{k-1}}{argmax}\; R^{\theta}(d)
\end{split}
\label{eq:fairness_select}
\end{equation}
In practice, Eq.~(\ref{eq:relevance_select}) and Eq.~(\ref{eq:fairness_select}) can be computed together by maintaining multiple priority queues $\mathbb{G}$, for each group separately. 

\subsubsection{Balance between Relevance and Fairness}
MMF implements a simple yet effective method to control the relevance and fairness of top-k rankings in dynamic LTR by adding a stochastic controller parameterized with $\lambda$.
Intuitively, the idea of the controller is to choose the final selected item $d_{\tau}^k$ between the document that maximizes ranking relevance $\bar{d_{\tau}^k}$ and the document maximizing marginal fairness $\Tilde{d_{\tau}^k}$ with probability $\lambda$.
Formally, we have
\begin{equation}
\begin{split}
d_{\tau}^k \sim (\lambda \Tilde{d_{\tau}^k} +(1-\lambda) \bar{d_{\tau}^k})
\end{split}
\label{eq:lambda_select}
\end{equation}
The greater $\lambda$ is, the fairer the ranking is and the less ranking relevance is. 
Different from the linear trade-off strategy used in other fairness algorithms \cite{singh2018fairness,morik2020controlling} that directly combines the fairness and relevance scores before the document selection process, we adopt a probabilistic strategy to strike the balance, which not only makes the algorithm more robust to the magnitude of estimated relevance and fairness scores, but also provides explicit functions to balance relevance and fairness in practical applications.

The overall MMF is shown in Algorithm~\ref{algo:controller}. 
At step $1$, we do initialization. 
At step $2$, a user enters the dynamic LTR system. 
At step $3-7$, we dynamically learn a LTR model $R^{\theta}$ or directly use the averaged inverse propensity weighted clicks $R^{IPS}$ to estimate the relevance of items.   
The LTR model has advantages over $R^{IPS}$ as it can estimate personalized relevance if features contain user information. At step $8-10$, we construct priority queues for each group with estimated relevance from step $3-7$. In practice, we could construct priority queues while estimating relevance at the time.
At step $12-21$, we select the item according to \Cref{eq:fairness_select,eq:relevance_select,eq:lambda_select}. At step $15$, $G_{ind}$ is a function to get group index of an item. At step $22$, we collect clicks in a real-world application or sample clicks according to Eq.\ref{eq:click} for our experiments. At step $23-27$, we update the relevance estimation accordingly.
Note that, for different time steps, users interact with the same item candidates at step $2-27$.

\begin{algorithm}
\SetAlgoLined
 initialize $\lambda$\ within [0,1], $k$, $\mathbf{c_t}\xleftarrow{}0$,
initialize $R^\theta$, $R^{IPS}(d)\xleftarrow{}0$\;
\For{each user (time step $\tau$)} {
  \uIf{Use\_LTR\_model}{
    Estimate relevance for all items with  model $R^\theta $ \;
  }
  \Else{ Estimate relevance for all items with $R^{IPS}(d)$ according to Eq.\ref{eq:average_relevance}\;}
\For{each group $G_i$ }{
construct priority queues $\mathbb{G}_i$ of size k for group $G_i$ with estimated relevance.}
ranking=[]\;

\For{each rank i }{
\uIf{ random>$\lambda$}{$d_i$ selected with Eq.\ref{eq:relevance_select}, and ranking.append($d_i$)\;$j=G_{ind}(d_i)$, index group id and assigin to j\;
$\mathbb{G}_j.pop()$
}
\Else{select $G_{\tau}^k$ according to Eq.\ref{eq:group_select} and assign $G_{\tau}^k$ to j\;
ranking.append($\mathbb{G}_{j}.pop()$)\;}
}
present ranking and collect user clicks $\mathbf{c_\tau}$, or sampling clicks  according to Eq.\ref{eq:click}\;
  \uIf{Use\_LTR\_model}{train $\widetilde{\mathcal{L}}^\tau(\theta)$ loss in Eq.\ref{eq:unbias}. \;
  }
  \Else{ update $R^{IPS}(d)$\ according to Eq.\ref{eq:average_relevance}\;}
}
\caption{MMF}

\label{algo:controller}
\end{algorithm}
\vspace{-0.2cm}

\subsection{Complexity Analysis}
To illustrate the efficiency of MMF, we conduct complexity analysis and use the state-of-the-art fairness algorithm, i.e., FairCo \cite{morik2020controlling}.
FairCo is one of the most efficient fairness algorithms for dynamic LTR.
It achieves fairness by dynamically adding perturbations to the ranking scores of documents based on the exposure of different groups.
As relevance is estimated separately, we only discuss the complexity of fairness controlling in FairCo and MMF.

\subsubsection{Time Complexity}
Let $k$ be the number of ranks we care about, $|\mathcal{G}|$ be the number of groups, and $n$ be the total number of documents to rank.
As FairCo needs to add score perturbations to all documents, it need to track the cumulative group exposure of all documents (i.e., $Exp\_cum_{\tau}^n(G_i)$ in Eq.~(\ref{eq:pair_disparity})) and do linear interpolation with $O( n)$ time.
It needs  $O(k\cdot log(n))$ to select top-k results from $n$ documents and thus has the overall time complexity as $O(k\cdot log(n) + n)$.
In contrast, MMF only tracks $Exp\_cum_{\tau}^k(G_i)$ for top-k results with $O(|\mathcal{G}|\cdot k)$ time.
Besides, it takes $O(|\mathcal{G}|\cdot k\cdot log(n))$ time to construct priority queue for each group, and takes $O(|\mathcal{G}|\times k)$ time to implement Eq.\ref{eq:group_select}.
The overall complexity of MMF is $O(k\cdot |\mathcal{G}|\cdot\big(1+log(n)\big))$. 
Therefore, MMF takes $O( (n-|\mathcal{G}|\cdot k)+k\cdot log (n) \cdot(1-|\mathcal{G}|))$ less time than FairCo.
Because the size of all documents is usually much more than the number of ranks and the number of groups we care (i.e., $n>>k$ and $n>>|\mathcal{G}|$) in most ranking applications, MMF outperforms FairCo in time.

\subsubsection{Space Complexity}
MMF needs to track of $Exp\_cum_{\tau}^k(G_i)$ in Eq.~(\ref{eq:pair_disparity}) for all positions in top-k rankings for the computation of marginal fairness, which has space complexity as $O(|\mathcal{G}|\cdot k)$.
FairCo only needs to track the overall cumulative exposure of all documents, which has space complexity as $O(n)$.
As the number of groups (e.g., race, gender) and the ranks we care are usually small, MMF outperforms FairCo in space.

\section{Experiments and results}
To evaluate our method, we conduct experiments on one dataset with simulated preference data (i.e., the News dataset \cite{harper2015movielens}) and one dataset with real-world preference data (i.e., the Movie dataset \cite{harper2015movielens}). 
All the experimental scripts and model implementations used on this paper is available online\footnote{\url{https://github.com/Taosheng-ty/Dynamic-Fairness.git}}.

\subsection{Simulated Preference Data}\label{sec:news_data}
In this paper, we create a simulated preference dataset with the news articles in the AdFrontes Media Bias dataset\footnote{\url{https://www.adfontesmedia.com/interactive-media-bias-chart/}}, which we refer to as the News dataset.
In this News dataset, each article contains a polarity value $\rho^d$ that has been rescaled to the range between -1 and 1 (i.e., left-leaning and right-leaning).
Following the methodology used by \citet{morik2020controlling}, we simulate a dynamic LTR problem on the News Dataset with simulated users and assign each user with a polarity preference drawn from a mixture of two guassian distirubtion and are also cliped to $[-1,1]$. 
\begin{equation}
\rho^{u_t}\sim clip_{[-1,1]}(p_{neg}\mathcal{N}(-0.5,0.2)+(1-p_{neg})\mathcal{N}(0.5,0.2))
\end{equation}
where $p_{neg}$ is the probability of the user to be left-leaning.
In addition, we simulate each user with an opensness parameter $o^{u_t}\sim \mathcal{U}(0.05,0.55)$, indicating on the breadth of interest outside their polarity. With sampled $u_t$, $\rho^{u_t}$, and articles' polarity value annotation $\rho^{d}$, we could synthesize a true binary relevance judgement following a Bernoulli distribution as 
\begin{equation}
\bm{r_t}(d)\sim Bernoulli\bigg[p=exp\bigg(\frac{-(\rho^{u_t}-\rho^d)^2}{2(o^{u_t})^2}\bigg) \bigg]
\end{equation}
In each experiment trial, we sample a set of 30 news articles $D$ to recommend to users.
To investigate ranking fairness, we group articles according to their polarity by assigning articles with $\rho^d\in[-1,0)$ to group $G_1$ and articles with $\rho^d\in[0,1]$ to group $G_2$.

We implement four baselines for comparison. The first one is the \textbf{Naive} method that ranks documents by the sum of their observed user clicks (i.e., $\bm{c_t}$).
The second one is a simple unbiased LTR algorithm \textbf{D\-ULTR(Glob)} that ranks documents by the unbiased relevance estimation $R^{IPS}(d)$ in Eq.~(\ref{eq:average_relevance}).
To show the effectiveness of MMF as a fairness algorithm, we also include two state-of-the-art fairness algorithms for dynamic LTR, which are \textbf{LinProg} \cite{singh2018fairness} and \textbf{FairCo} \cite{morik2020controlling}.
Proposed by \citet{singh2018fairness}, LinProg considers group fairness as optimization constraints and implement a linear programming algorithm to maximize ranking fairness based on the relevance estimated by $R^{IPS}(d)$. 
In theory, LinProg would produce the best ranked lists in terms of merit-based fairness.
FairCo achieves group fairness by dynamically adding perturbations to the ranking scores of documents based on the exposure of different groups.
It has been proven to be effective in optimizing the fairness of dynamic LTR and much more efficient than LinProg. 

For all models in the experiments on the News dataset, we use $R^{IPS}(d)$ to estimate document relevance directly as all users are simulated. 
This means $Use\_LTR\_model$ in Algo.\ref{algo:controller} is set as False. 
LinProg and FairCo use a hyper-parameter to reduce unfairness of exposure over all documents.
For simplicity, we set it as 0.1 for LinProg and 0.01 for FairCo as these were the tuned hyper-parameters in \citet{morik2020controlling}.
For MMF, we tune $\lambda$ from 0 to 1 and show the corresponding results in Table~\ref{tab:dataset1} and Figure~\ref{fig:comparison_NDCG_unfairness_at_10}.
Besides, We follow the experiment setting in  \cite{morik2020controlling} and adopt the discount function of NDCG as the user's user examination probability for each position for this paper in simulation,
\begin{equation}
    p_i=\big(\frac{1}{log2(1+i)} \big)
\end{equation}
where $p_i$ indicates examination probability at rank $i$.
\subsection{Results with Simulated Preference Data}
\begin{table*}[t]
	\caption{Comparison of MMF with different baselines on News and Movie. Significant improvements or degradations with respect to the performance of FairCo are indicated with $+/-$ in the paired t-test with p-value  $p \leq 0.05$. The best performance of fair algorithms in each column is highlighted in boldface.
	}
	\vspace{-0.2cm}
	\begin{subtable}[t]{1.0\textwidth}
		\centering
		\small	
		\caption{Performance of learning-to-rank algorithms on News data.}
		\scalebox{1.0}{
			\begin{tabular}{ p{0.10\textwidth}| c || c | c | c  | c || c |c| c  | c     } 
				\hline
				\hline
				\multicolumn{2}{c||}{} &  NDCG@3 & NDCG@5 & NDCG@10  & NDCG@all   & Unfairness@3  & Unf.@5  & Unf.@10 & Unf.@all\\ \hline
				\hline
				\multirow{2}{0.10\textwidth}{Unfair Algorithms} 
				& Naive & 0.418$^-$& 0.430$^-$& 0.470$^-$& 0.697$^-$& 0.220$^-$& 0.268$^-$& 0.293$^-$& 0.174$^-$\\ \cline{2-10}
				& D\_ULTR(Glob) & 0.438$^+$& 0.448$^+$& 0.490$^+$& 0.708& 0.188$^-$& 0.218$^-$& 0.242$^-$& 0.136$^-$\\ 
				\hline
				\hline
				\multirow{2}{0.10\textwidth}{Fair Algorithms}	&FairCo  & 0.434& 0.443& 0.483& 0.705& 0.036& 0.037& 0.049& 0.015 \\ \cline{2-10}
				& LinProg& 0.433& 0.439& 0.462& 0.694& 0.043& 0.051& 0.065& \textbf{0.010}\\
				\cline{2-10}
			 
				
				
				
				
				
				& MMF($\lambda=0.6$)  & \textbf{0.436}& \textbf{0.447}$^+$& \textbf{0.488}$^+$& \textbf{0.708} & \textbf{0.004}$^+$& \textbf{0.005}$^+$& \textbf{0.007}$^+$& 0.020
				
				
				\\ \hline
				\hline
			\end{tabular}
		}
		
		\label{tab:dataset1}
	\end{subtable}
		\begin{subtable}[t]{1.0\textwidth}
		\centering
		\small	
		\caption{Performance of learning-to-rank algorithms on Movie data.}
		\scalebox{1.0}{
			\begin{tabular}{ p{0.10\textwidth}| c || c | c | c | c || c |c| c |  c     } 
				\hline
				\hline
				\multicolumn{2}{c||}{} &  NDCG@3 & NDCG@5 & NDCG@10  & NDCG@all  & Unfairness@3  & Unf.@5  & Unf.@10 & Unf.@all\\ \hline
				\hline
				\multirow{3}{0.10\textwidth}{Unfair algorithm} 
				& Naive & 0.633$^-$& 0.638$^-$& 0.652$^-$& 0.761$^-$& 0.166$^-$& 0.168& 0.111$^+$& 0.265$^-$\\ \cline{2-10}
				& D\_ULTR(Glob) & 0.671$^-$& 0.665$^-$& 0.672$^-$& 0.775$^-$& 0.102$^+$& 0.105$^+$& 0.066$^+$& 0.243$^-$\\ 
				\cline{2-10}
				& D\_ULTR & 0.827$^+$& 0.819$^+$& 0.816$^+$& 0.871$^+$& 0.035$^+$& 0.036$^+$& 0.042$^+$& 0.232$^-$\\

				\hline
				\hline
				\multirow{2}{0.10\textwidth}{Fair algorithm}	&FairCo  & \textbf{0.814}& 0.802& 0.791& 0.850& 0.123& 0.162& 0.234& \textbf{0.021}\\
				\cline{2-10}
			 
				
				

				& MMF($\lambda=0.1$)
				& 0.810& \textbf{0.804}& \textbf{0.802}$^+$& \textbf{0.863}$^+$& \textbf{0.008}$^+$& \textbf{0.011}$^+$& \textbf{0.016}$^+$& 0.312$^-$
				
				
				
				
				
				\\ \hline
				\hline
			\end{tabular}
		}
		
		\label{tab:dataset2}
	\end{subtable}
	\label{tab:dataset}
\vspace{-0.5cm}
\end{table*}

\subsubsection{Do the unbiased estimates converge to the true relevance no matter fairness controlling is applied or not? }

\begin{figure}
    \centering
    \vspace{-0.4cm}
    \includegraphics[scale=0.52]{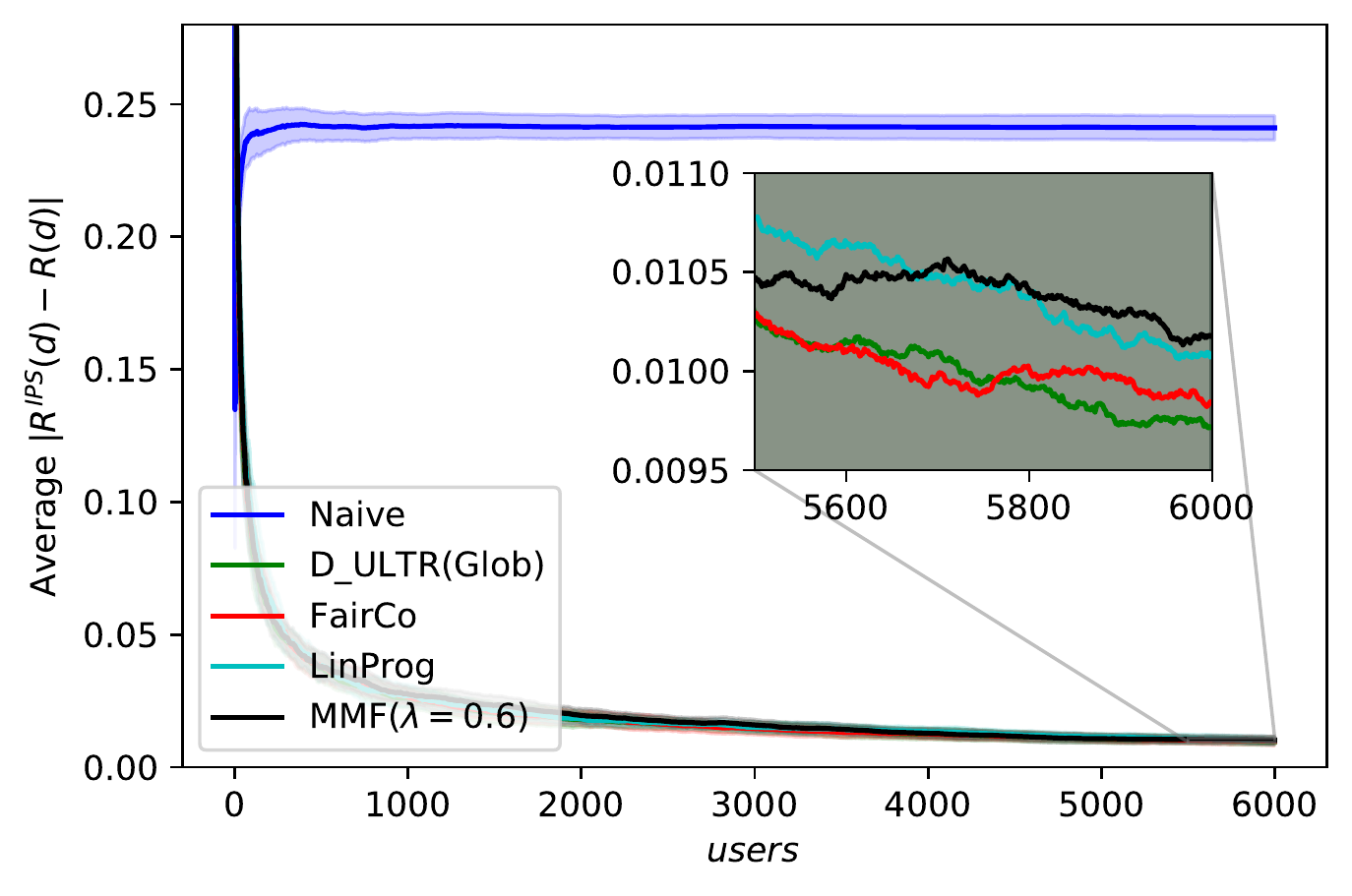}
    \vspace{-0.4cm}
    \caption{The absolute difference between estimated global relevance and true global relevance on News (20 trials).}
    \label{fig:relevance_converge}
\end{figure}
Fig.~\ref{fig:relevance_converge} shows the absolute difference between estimated global relevance and true global relevance defined in Eq.~(\ref{eq:average_relevance}) after applying different ranking algorithm. The error of all algorithm with IPS weighting (Linprog,FairCo,D\_ULTR(Glob) and MMF($\lambda=0.6$)) gradually approach zero, while the error of Naive algorithm keeps around 0.25. There are no significant difference of convergence between IPS weighted algorithms which means we could exclude the influence of relevance estimation when comparing ranking and fairness controlling for IPS weighted algorithms. The result verifies that IPS weighting help to get an unbiased estimation of the true expected relevance for each news articles no matter what ranking algorithm is used and thus estimated relevance can be used to maximize ranking and fairness utility. 

\subsubsection{Can $MMF$ effectively reduce unfairness while maintaining good ranking performance? }
\begin{figure}
    \centering
    \vspace{-0.4cm}
    \includegraphics[scale=0.52]{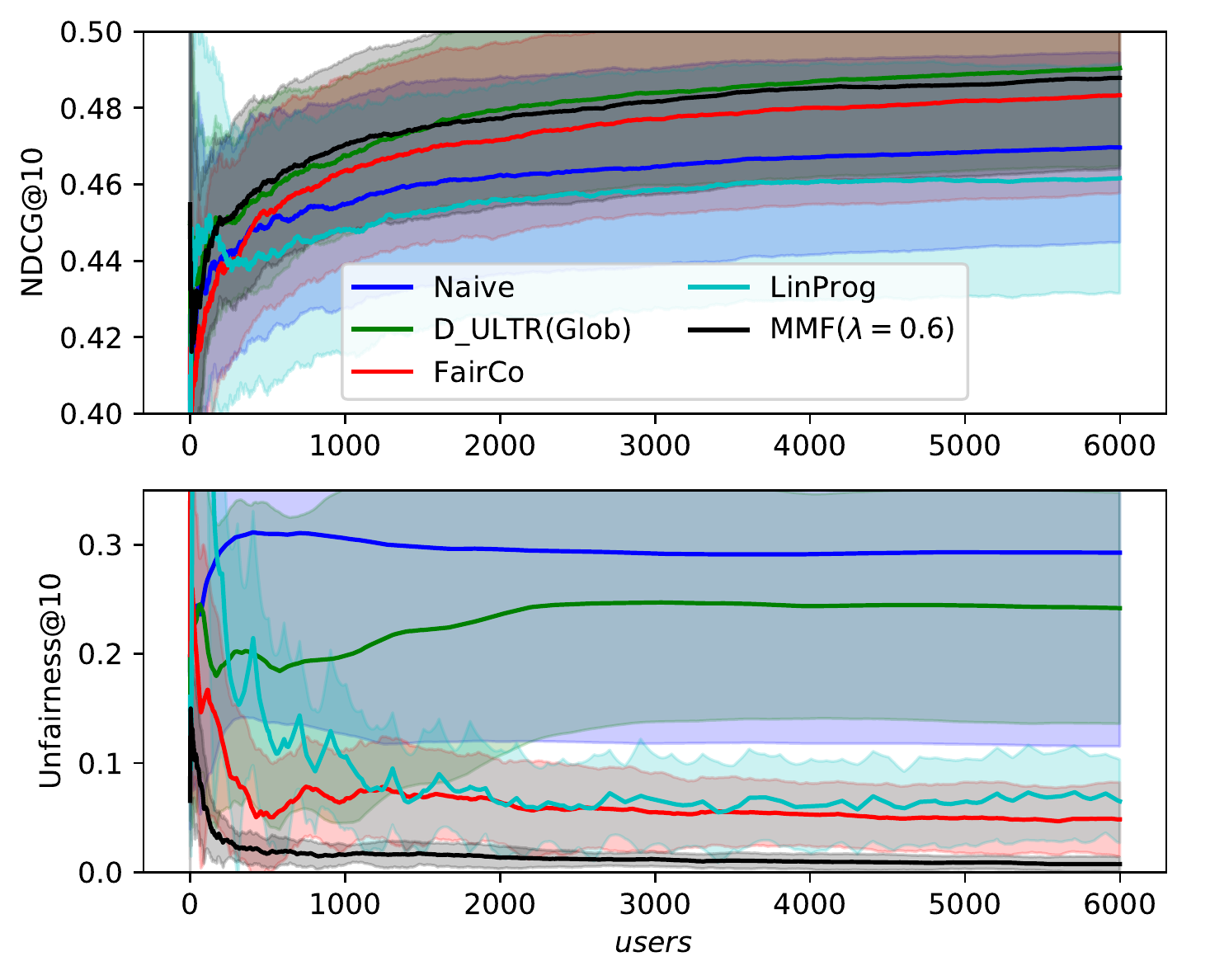}
    \vspace{-0.4cm}
    \caption{Convergence of NDCG$@10$ and Unfairness$@10$ as the number of users increases on News. (20 trials)}
    \label{fig:convergence_at_5}
    \vspace{-0.4cm}
\end{figure}

Fig.\ref{fig:convergence_at_5} shows the convergence NDCG$@10$ and unfairness$@10$  for Naive, D-ULTR(Glob), FairCo, Linprog and MMF~($\lambda=0.6$). Among them, Naive shows the highest unfairness, while NDCG remains low as user interaction increases. D-ULTR(Glob) achieve the best NDCG while its unfairness is worse than all other algorithms except Naive. Our algorithm MMF manages to substantially reduce unfairness while achieve similar NDCG performance with D\_ULTR(Glob). For the other two algorithm, LinProg and FairCo both show inferior performance than MMF in terms of NDCG and unfairness.

\subsubsection{How $MMF$ performs at different prefixes of a ranking? }

\begin{figure}
    \centering
    \vspace{-0.4cm}
    \includegraphics[scale=0.52]{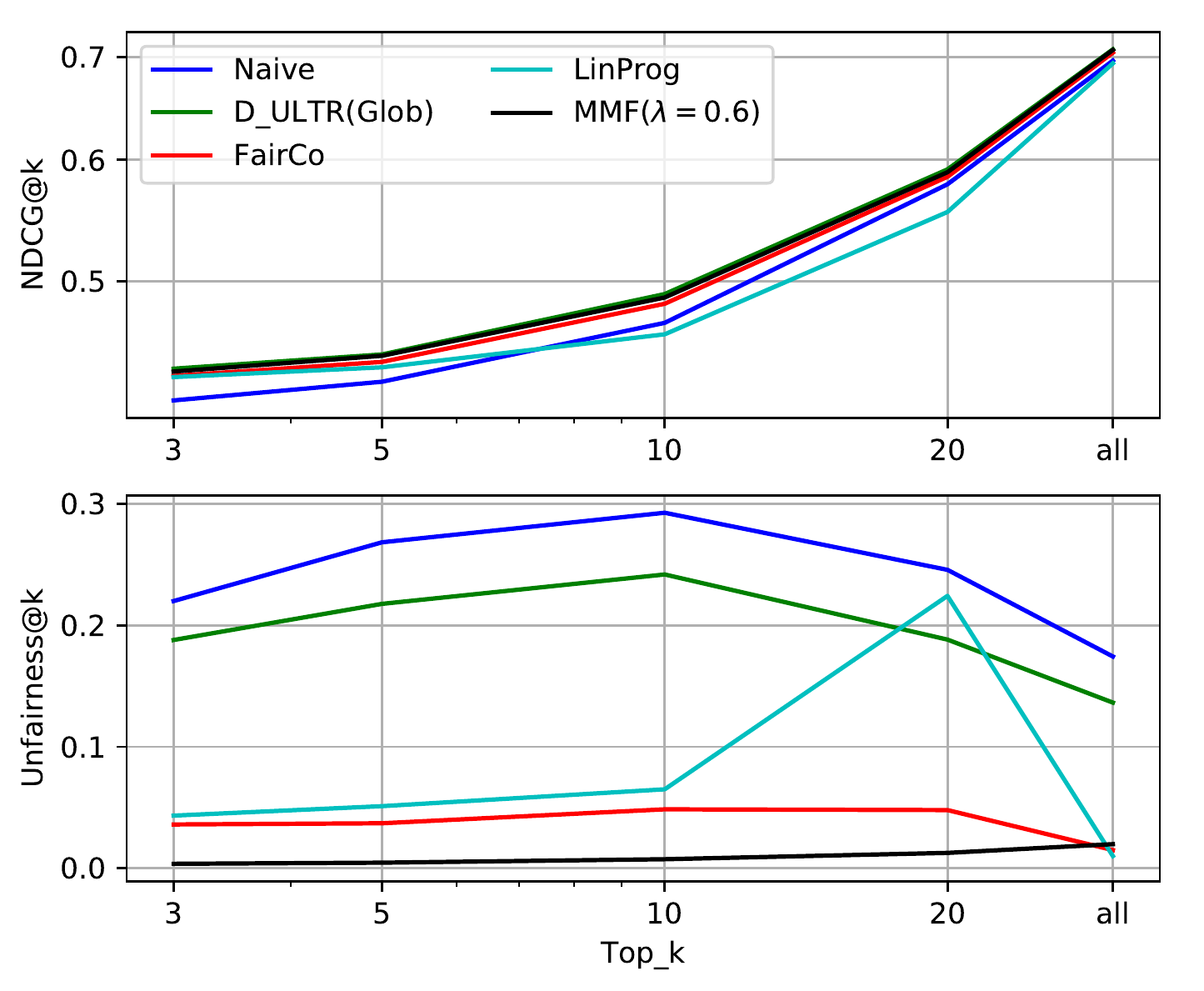}
    \vspace{-0.4cm}
    \caption{Performance of NDCG(top) and Unfairness (bottom) for different prefixes on News (20 trials).}
    \vspace{-0.4cm}
    \label{fig:comparison_top_k}
\end{figure}
Fig.\ref{fig:comparison_top_k} shows the performance NDCG and unfairness at different prefixes for Naive, D-ULTR(Glob), FairCo,Linprog and MMF($\lambda=0.6$ after 6000 users interactions. More details are shown in Table \ref{tab:dataset1}.
In terms of NDCG performance, Naive shows worse performance than IPS weighting algorithms, especially in top ranks, which make sense since richer-get-richer dynamics can be more severe in top ranks. 
In terms of fairness, among all algorithms, Naive shows the worst performance almost for top ranks. 
Among the fairness controlling algorithms, Linprog and FairCo show superior performance on overall fairness (i.e., unfairness$@$all). As is shown in Table.\ref{tab:dataset1}, compared to Linprog and FairCo, MMF performs much better on top ranks while showing a slight compromise on overall unfairness.
This may attribute to that MMF try to mitigate unfairness for each prefix, while Linprog and FairCo always consider all documents. We think this is the key advantage of our algorithm compared with Linprog and FairCo since top ranks are relatively important for rankings, and many users would leave after viewing top results. 
\subsubsection{How does $\lambda$  control trade-off between ranking performance and unfairness for $MMF$?}
\begin{figure}
    \centering
    \includegraphics[scale=0.52]{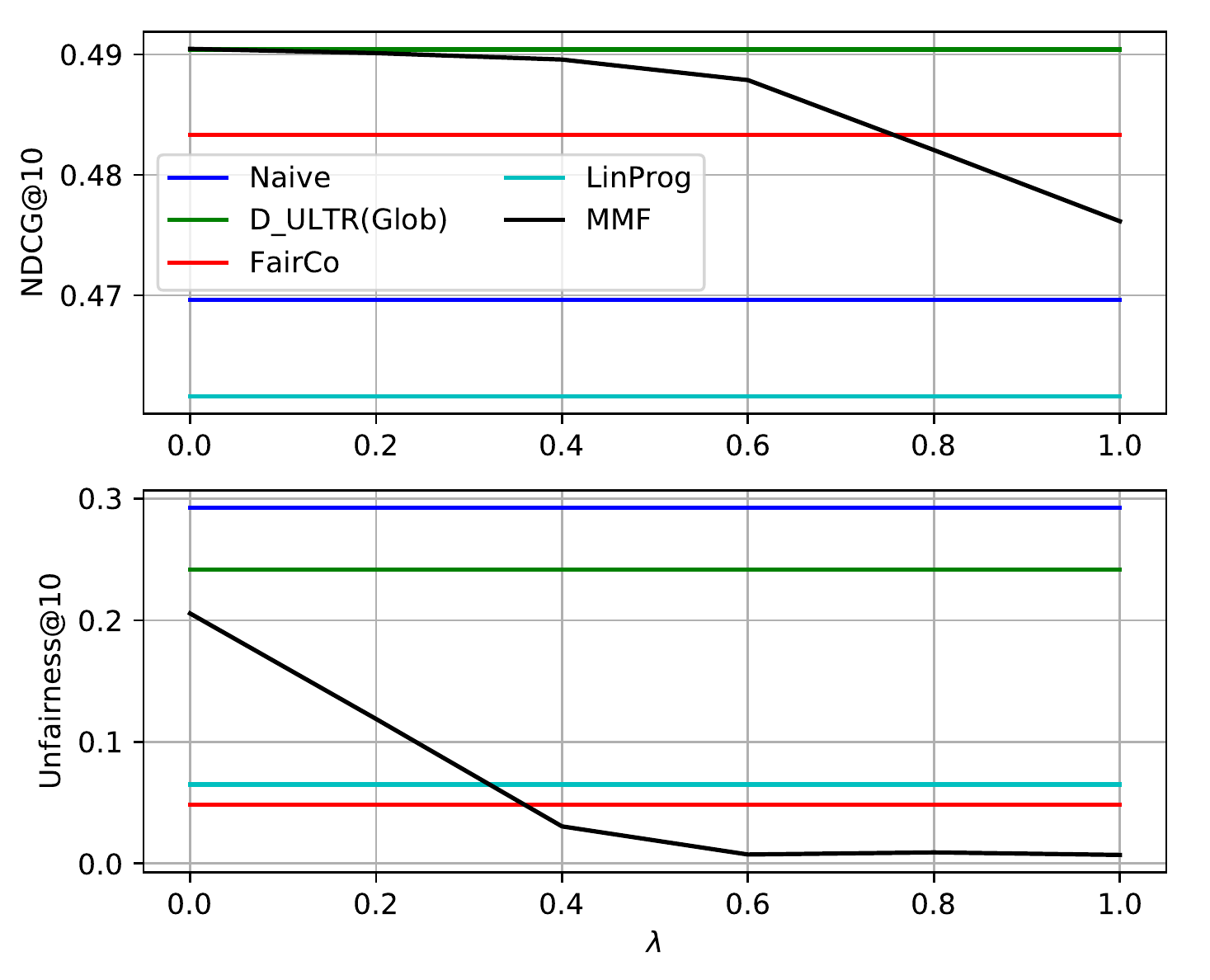}
    \vspace{-0.4cm}
    \caption{Performance of NDCG$@10$(top) and Unfairness$@10$ (bottom) of MMF with different $\lambda$ on News (20 trials).}
    \label{fig:comparison_NDCG_unfairness_at_10}
    \vspace{-0.4cm}
\end{figure}
Fig.~\ref{fig:comparison_NDCG_unfairness_at_10} shows the performance NDCG$@10$ and unfairness$@10$ for  MMF with different $\lambda$ after 6000 users interactions.  
Note that we only show LinProg and FairCo with the tuned hyper-parameters as shown in  \cite{morik2020controlling}. Thus their performances are constants. 
As we can see, the use of hyper-parameter $\lambda$ in MMF enables us to explicitly control the trade-off between relevance and fairness.
When $\lambda=0.0$, MMF degenerates to D\_ULTR(Glob), thus it has similar unfairness with D\_ULTR(Glob), greater than Linprog and FairCo.
With the increasing of $\lambda$, unfairness gradually decreases and become less than Linprog and FairCo on top results. 
From the figure, we can see that choosing $\lambda$ from 0.4 to 0.7 for MMF can achieve better performance than the baseline algorithms on both relevance and fairness.

\subsection{Real-World Preference Data}
To evaluate our method on a real-world preference data, we use the ML\-20M dataset, which we refer to as the Movie dataset.
Following the prepossessing method in  \cite{morik2020controlling}, we select five production companies with the most movies in the dataset(MGM, Warner Bros, Paramount, 20th Century Fox, Columbia). We aim to ensure fairness of exposure for films from the five production companies, which means movies from the same company belong to the same group. 
A set of 300 most rated movies by those production companies are selected. Then the 100 movies with the highest standard deviation in the rating across users are selected. For users, we select $10^4$ users who have rated the most number of the chosen 100 movies. 
Finally, we get a partially filled rating matrix with $10^4$ users and 100 movies. 
We use an off-the-shelf matrix factorization algorithm\footnote{Surprise library (\url{http://surpriselib.com/}) for SVD with biased=False and D=50} to fill the missing entries. We then normalize the rating to $[0,1]$ by applying a Sigmoid function centered at rating $b=3$ with slope $a=10$. Thus, it can serve relevance probabilities where higher star ratings correspond to higher likelihoods of positive feedback, which is used to generate clicks. We use the user embedding from the matrix factorization model as user features $x_t$. 
We keep the dimension of user features to 50.
At each time step $t$, we sample a user $x_t$ and the ranking algorithm presents a ranking of the 100 movies.
For the user personal relevance estimation model $R^\theta$ used by FairCo and D\-ULTR, and our method, we use a one hidden-layer neural network that consists of $D=50$ input nodes, which corresponds to user feature dimension, then fully connected to 64 nodes in the hidden layer with RELU activation, which is then connected 100 output nodes with Sigmoid to output the predicted relevance probability for the 100 selected movies.

Besides the baselines discussed in Section~\ref{sec:news_data}, we also include a new baseline in the experiments on the Movie dataset, which is \textbf{D-ULTR}.
D-ULTR conducts unbiased learning to train the LTR model with click data.
Different from D-ULTR(Glob) that directly ranks documents with $R^{IPS}(d)$, D-ULTR can model personalized relevance by taking user features into account when building the LTR model. 
Thus, we expect D-ULTR to perform much better than D-ULTR(Glob) on the Movie dataset with real user preferences.
Similarly, we train LTR models with user features for both FairCo and MMF.
Note that we exclude LinProg on the Movie dataset as it is designed to work with global relevance and is too computationally expensive for rankings with a large number of documents.

\subsection{Results with Real-World Preference Data}
\subsubsection{Can $MMF$ effectively reduce unfairness while maintaining good ranking performance? }
\begin{figure}
    \centering
    \includegraphics[scale=0.52]{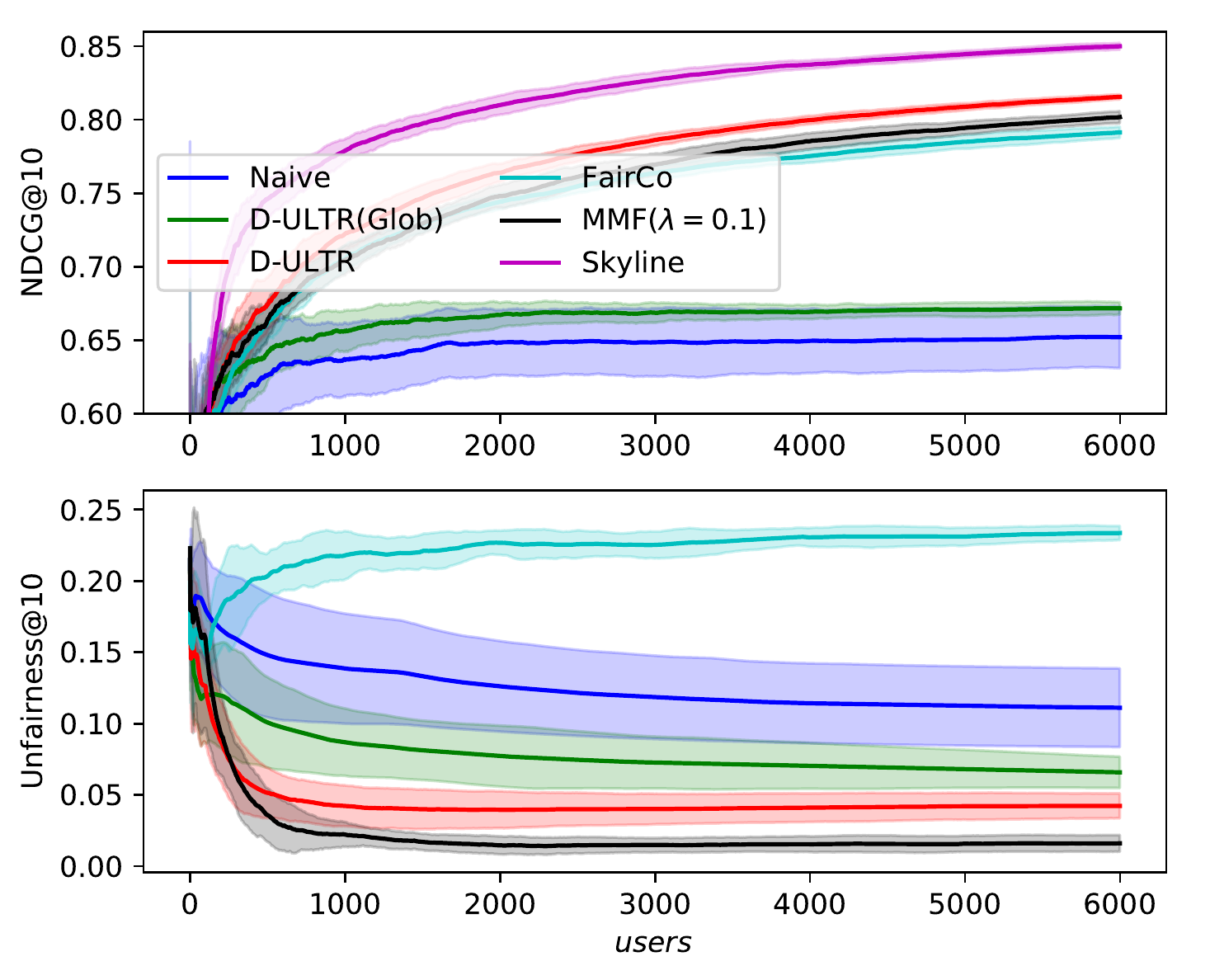}
    \vspace{-0.4cm}
    \caption{Convergence of NDCG$@10$ and Unfairness$@10$ as the number of users increases on Movie (5 trials).}
    \label{fig:d2:convergence_10}
    \vspace{-0.6cm}
\end{figure}
We show the performance of ranking relevance and fairness for all algorithms on the Movie dataset in Fig.~\ref{fig:d2:convergence_10}
For references, we plot a \textbf{Skyline} model that trains the LTR model with ground truth relevance judgements and ranks documents via the output estimated relevance from the LTR model. 
Firstly, as is shown in Fig.~\ref{fig:d2:convergence_10}, personalization do help to reach better ranking performance just as  \cite{morik2020controlling} already reported. 
Ranking algorithms (D\-ULTR, FairCo, MMF, skyline) relying on personalized relevance show superior ranking performance than algorithms like Naive and D\-ULTR(Glob), where no personalization is used.
Ranking algorithms involving IPS and personalization can approach the skyline as more user interactions are available, which again verifies the effectiveness of IPS and personalization.

After we show the effectiveness of personalization and IPS, we now take a look at the overall performance shwon in Table \ref{tab:dataset2}. Naive and D\-ULTR(Glob) show the worst performance in terms of ranking and fairness since no personalization or fairness controlling involved. Compared to FairCo, our method MMF($\lambda=0.1$) show a slight compromise on NDCG@3 while achieving better performance on NDCG@5, NDCG@10, NDCG@all, and much better performance on fairness metrics on top results. 
The fairness controlling of FairCo is not effective in optimizing fairness on top-k ranks. 
We think the reason might be that FairCo focuses on mitigating overall unfairness instead of top ranks, which means it ignores fairness of top ranks. Such phenomenon can be clearly shown in Fig.~\ref{fig:d2:top_k}.

\subsubsection{How does $MMF$ perform at different prefix of a ranking?  }
\begin{figure}
    \centering
    \includegraphics[scale=0.52]{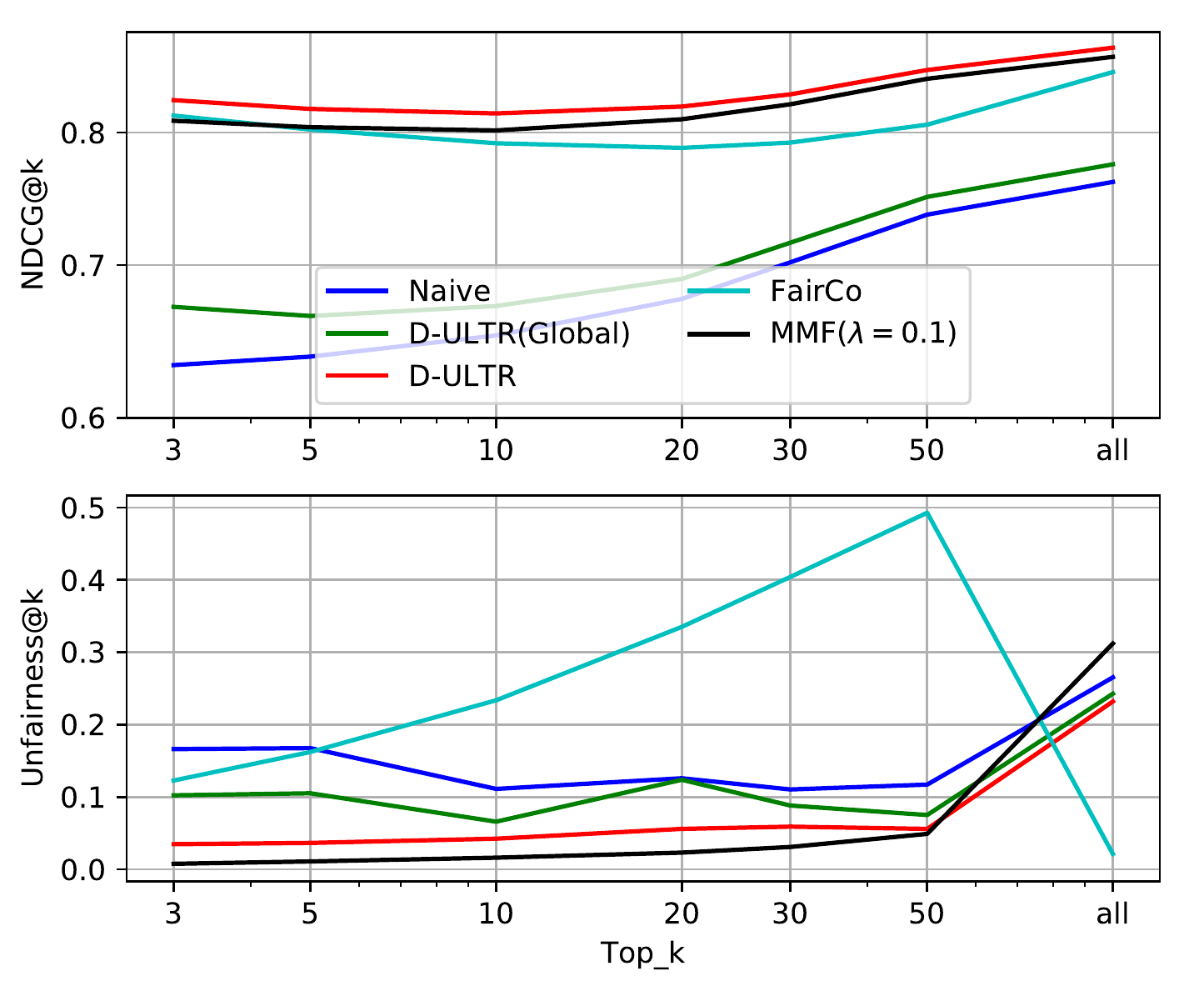}
    \vspace{-0.1cm}
    \caption{Performance of NDCG(top) and Unfairness (bottom) of MMF with different  prefixes on Movie (5 trials).}
    \label{fig:d2:top_k}
\end{figure}
Fig.\ref{fig:d2:top_k} shows the performance NDCG and unfairness at different prefixes for Naive, D-ULTR(Glob),ULTR, FairCo,MMF ($\lambda=0.1$) after 6000 users interactions.
As we can see, MMF achieves similar performance with D-ULTR in terms of relevance ranking while maintaining superior performance on ranking fairness on top results.
Specifically, it significantly outperforms FairCo on both relevance and fairness for $k$ from 3 to 50.
While FairCo has excellent performance on unfairness@all, it sacrifices both relevance and fairness performance on top results significantly.
It even shows more unfairness on top results than unfair algorithms such as D-ULTR.

\subsubsection{How does $\lambda$ control trade-off of ranking performance and mitigating unfairness for $MMF$}
\begin{figure}
    \centering
       \vspace{-0.1cm}
    \includegraphics[scale=0.52]{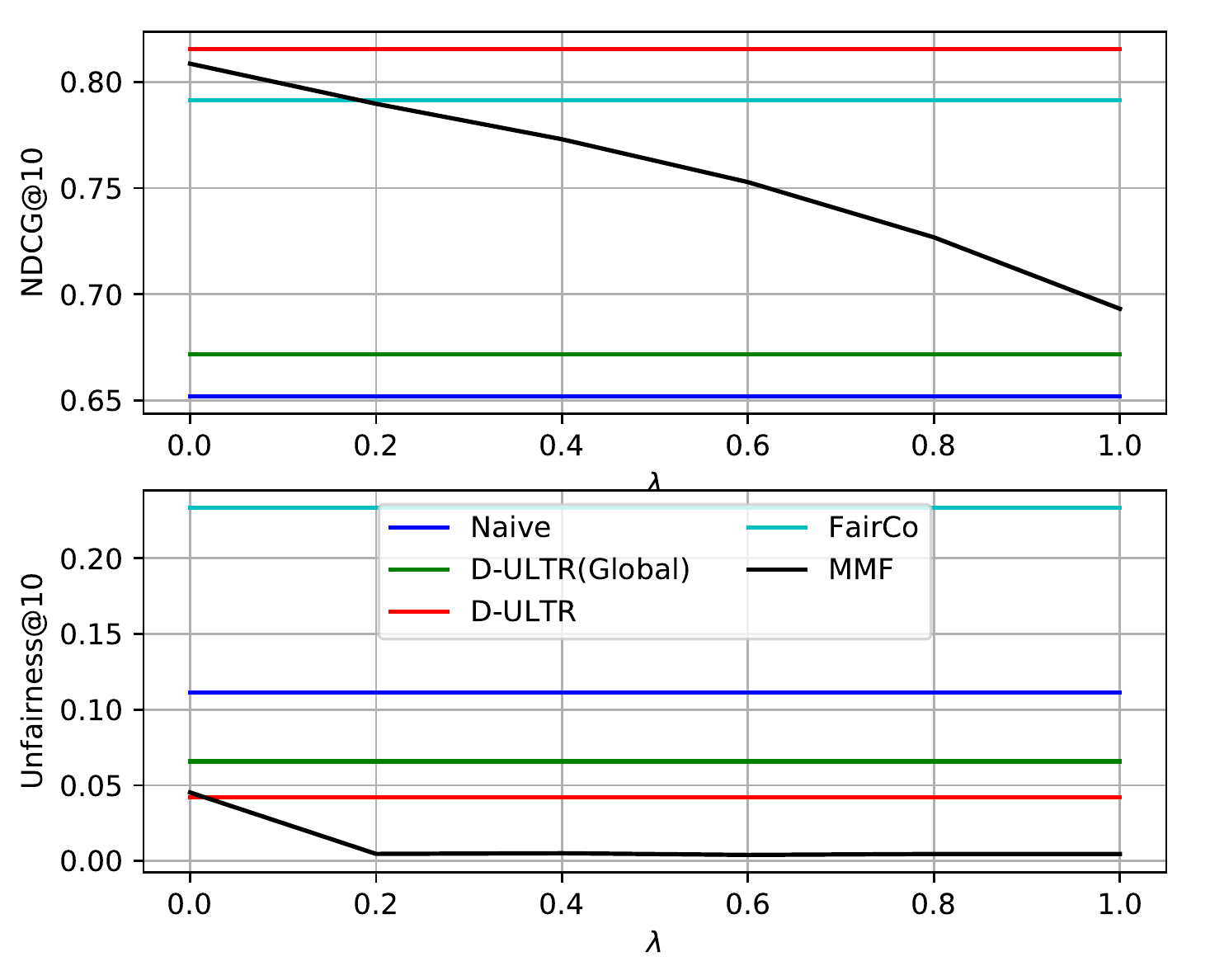}
    \vspace{-0.4cm}
    \caption{Performance of NDCG$@10$(top) and Unfairness$@10$ (bottom) of MMF with different $\lambda$ on Movie (5 trials).}
    \label{fig:d2:comparison_NDCG_unfairness_at_10}
    \vspace{-0.6cm}
\end{figure}
Fig.~\ref{fig:d2:comparison_NDCG_unfairness_at_10} shows the performance NDCG$@10$ and unfairness$@10$ for  MMF with different $\lambda$ after 6000 users interactions.  
Again, we report FairCo with the best parameter settings here. 
As $\lambda$ gradually increases, a trade-off can be seen from the growing of fairness (negative unfairness) and decreasing of ranking performance. 
And we may choose $\lambda$ to be between 0.0 and 0.2, between which MMF have better ranking performance as well as less unfairness. The optimal $\lambda$ for Movie data is smaller than for News data. We think it's because of their different relevance distribution. Documents in the News dataset often have similar relevance. In such situation, we should pay more attention to fairness (greater $\lambda$) for News dataset since relevance of different documents are close to each other. 

\section{CONCLUSION AND FUTURE WORK}
In this work, we propose a concept of marginal fairness and a Maximal Marginal Fairness (MMF) algorithm for balancing the relevance and fairness of top-k results in dynamic learning to rank.
We develop a metric to measure the group fairness of exposure in the top-k results of each ranked list and show that most existing state-of-the-art methods for ranking fairness focus more on the overall fairness of document exposure while compromising a lot in top ranks of each ranked list.
In contrast, our proposed MMF algorithm explicitly maximizes the marginal fairness of top-k rankings and can produce better rankings than the state-of-the-art fairness algorithms in both top-k relevance and top-k fairness.
In the future, we will further explore the possibility of extending MMF for more general ranking scenarios or construct new LTR models that integrate the optimization of relevance and fairness from the bottom of model design.
\begin{acks}
\vspace{-0.1cm}
This work was supported in part by the School of Computing, University of Utah. Any opinions, findings and conclusions or recommendations expressed in this material are those of the authors and do not necessarily reflect those of the sponsor.
\end{acks}
\newpage

\bibliographystyle{ACM-Reference-Format}
\bibliography{mybib}
\end{document}